\documentclass[12pt]{iopart}

\usepackage{iopams}
\usepackage{setstack}
\usepackage[T1]{fontenc}
\usepackage[utf8]{inputenc}
\usepackage{graphicx}
\usepackage{units}

\begin{document}

\title{Investigation of magnetic noise in Advanced Virgo}

\author{A Cirone$^{1,2}$, I Fiori$^3$, F Paoletti$^4$, M M Perez$^5$, A R Rodríguez$^5$, B L Swinkels$^6$, A M Vazquez$^5$, G Gemme$^1$, A Chincarini$^1$}

\address{$^1$ INFN, Sezione di Genova, I-16146 Genova, Italy}
\address{$^2$ Dipartimento di Fisica, Università degli Studi di Genova, I-16146 Genova, Italy}
\address{$^3$ European Gravitational Observatory (EGO), I-56021 Cascina, Pisa, Italy}
\address{$^4$ INFN, Sezione di Pisa, I-56127 Pisa, Italy}
\address{$^5$ IFAE, Barcelona Institute of Science and Technology, Barcelona, and ICREA; Spain}
\address{$^6$ Nikhef, Science Park 105, 1098 XG Amsterdam, The Netherlands}

\ead{alessio.cirone@ge.infn.it}

\begin{abstract}
The Advanced Virgo (AdV) sensitivity might be influenced by the effects of environmental noise, in particular magnetic noise (MN). In order to show the impact on the gravitational-wave strain signal h(t) and on the AdV sensitivity, we must understand the coupling between the environmental magnetic activity and the strain. The relationship between the environmental noise - measured by a physical environment monitor (PEM) - and h(t) is investigated using injection studies, where an intentional stimulus is introduced and the responses of both PEM sensors and the instrument are analyzed. We also present the most outstanding measurements and results obtained from both the characterization and the mitigation studies of the environmental MN. Results show that MN does not affect AdV sensitivity up to $\approx 100$ Mpc in BNS range.
\end{abstract}


\submitto{\CQG}

\maketitle

\section{Introduction}
AdV \cite{advVirgo,tdr} is a major upgrade of the Virgo/Virgo+ experiment, which significantly enhanced sensitivity to gravitational waves. The detector is a laser interferometer which consists of two $3\,\rm km$ long Fabry-Perot resonant cavities of a Michelson interferometer, fitted with four highly reflective 42 kg suspended mirror Test Masses (TM). The overall improvement of the detector allowed to reach a stable sensitivity of $\approx 50$ Mpc, in terms of Binary Neutron Star Inspiral Range, during the third scientific data taking.

The AdV sensitivity curve is shaped by several independent noises, which pose a challenge to the ongoing upgrading efforts. Among these disturbances, we focused on the magnetic noise (MN). In a recent paper \cite{cirone_magnetic_2018}, we aimed at finding the contribution of the MN to the entire noise budget of AdV. This was done by simulating the electromagnetic (EM) response of the payload to a slowly time-varying magnetic field with the use of Finite Element Analysis (FEA). We want to further investigate it by taking on-site surveys in order to improve our understanding of the interactions of MN with the whole AdV structure.

The main coupling channel between the outside EM environment and the AdV output data stream has been identified in the coil-magnet pairs which act on the TMs and the upper stages of the mirrors attenuation system for fine positioning. The reader can find a schematic illustration of the TM assembly and suspension system in the supplementary material (Figure S1, courtesy of \cite{cirone_magnetic_2018}).

The voice-coil actuators are very sensitive to the presence of the environmental magnetic field, whose fluctuations couple directly to the magnets and induce a mirror displacement noise. Even though the magnets are glued on the surface of the mirrors following an anti-parallel cross configuration (to compensate the effect of a uniform external magnetic field), in practice, the external field interacts with all the metallic structures surrounding the mirrors. These, in turns, cause eddy currents and warp the magnetic field with non-uniform gradients.

In this work, we did an extended magnetic field measurement campaign inside the main AdV buildings, in order to isolate the most powerful localized magnetic noise sources. This activity served as a preparation for the actual experimental measure of the transfer function between the ambient MN and the interferometer output. This measurement campaign was performed through a series of artificial noise injections. Then we compared results with two other possible ways to evaluate the relationship between the environmental noise and the gravitational wave strain signal h(t): the FE simulations done in 2016 \cite{cirone_magnetic_2018} and a long time period coherence-based approach. Finally, we will propose a feasible mitigation strategy in case we would need it for a later detector improvement.

\section{Magnetic characterization of the site}

\begin{figure*}[ht]
\centering
\includegraphics[width=2.5in]{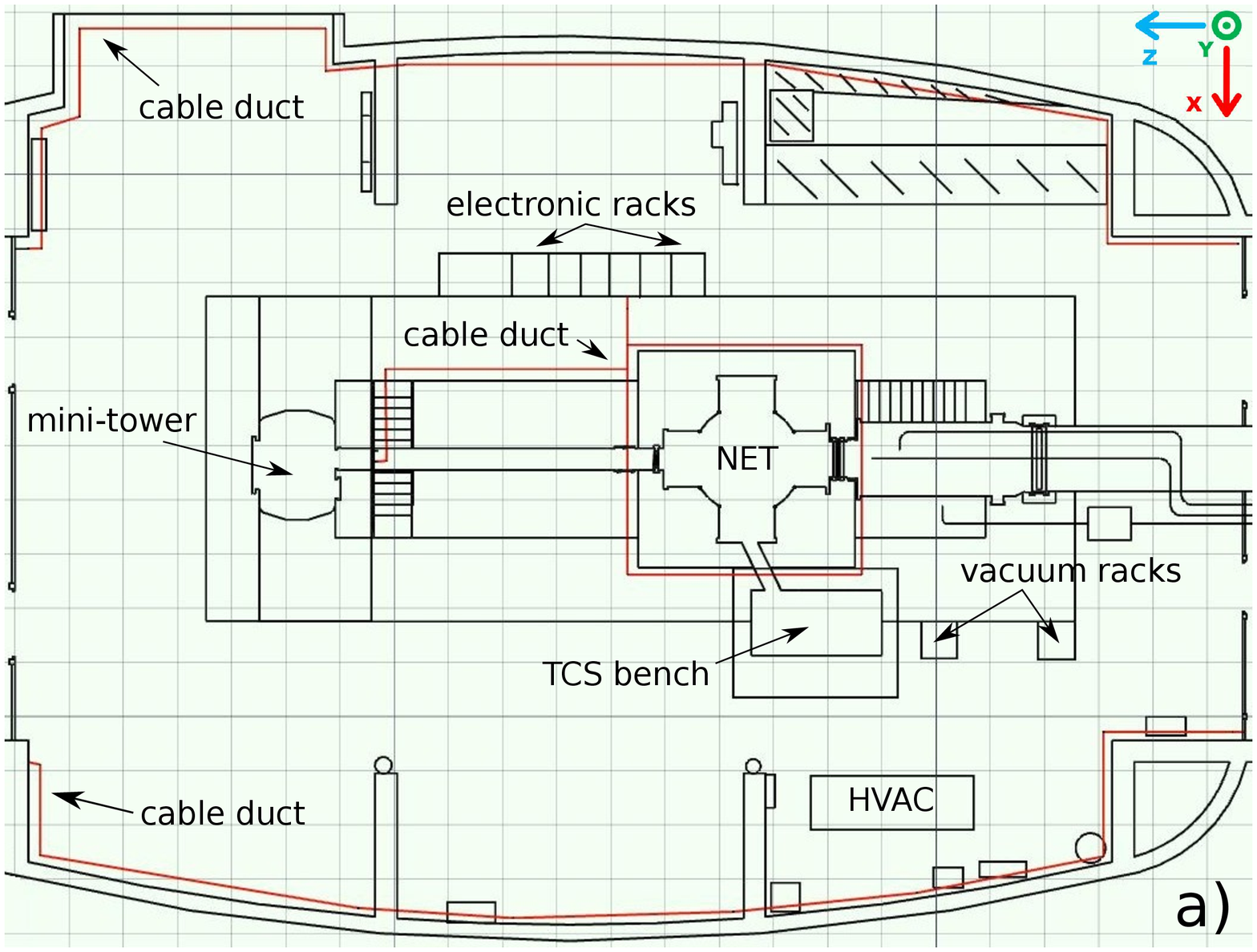}
\includegraphics[width=3.4in]{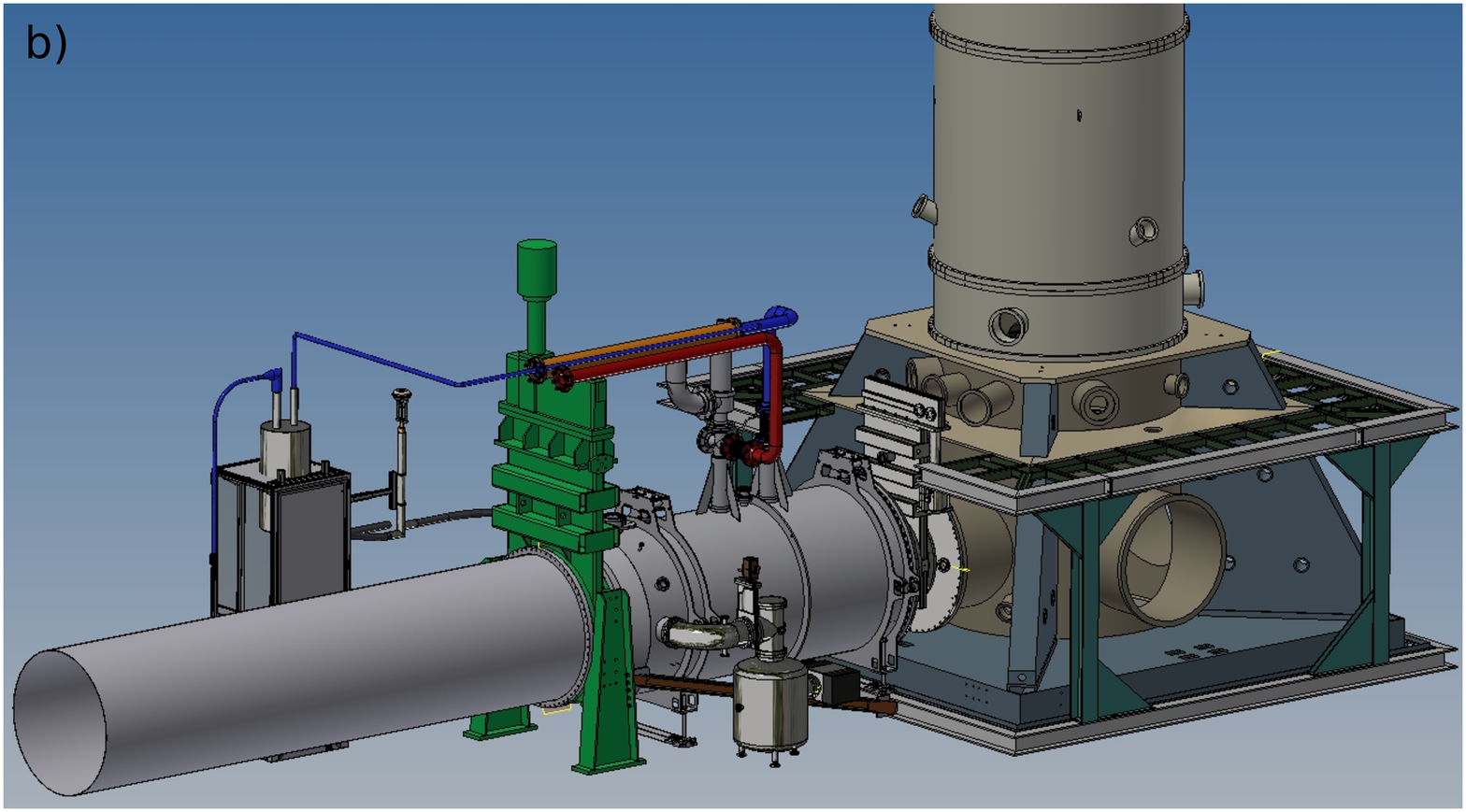}
\caption{a) Top-view drawing of the "North-End Building" experimental area and the main instrumentation. TCS, HVAC and NET stand respectively for "Thermal Compensation System", "Heating, Ventilation and Air Conditioning" and "North-End Tower". b) 3D CAD picture of the relevant infrastructure, which mainly consists of the vacuum-tight enclosure of the interferometer components.}
\label{fig:nebcad} 
\end{figure*}

In this first section, we discuss the extended magnetic measure campaign performed inside the three main AdV buildings: the Central Building (CEB), hosting the two input TMs, the West-End Building (WEB) and the North-End Building (NEB), hosting the two end TMs. In figure \ref{fig:nebcad}a-b we see the blueprint of NEB, where we recognize the vacuum chamber in the middle (also called North-End Tower, NET) hosting the north-end mirror and its seismic attenuation system, and all other ancillary devices. Depending on the availability of the area and the type of measure we want to perform, we choose either a terminal or the central building as a proxy for all the other AdV infrastructures, knowing that the CEB hosts far more equipment than the others.

\subsection{Typical magnetic sources}
\label{sec:mag_sources}

The sources of ambient MN are distributed throughout the experimental hall and include electronic boards, pumps, motors, lights, electrical power circuits and, in principle, any wire where current flows. Furthermore, eddy currents induced by the conductive nature of materials are not easy to measure. Because of eddy currents, external magnetic fields are spatially distorted and changed in magnitude by the presence of conductive parts.

\begin{figure*}[ht]
\centering
\includegraphics[width=4in]{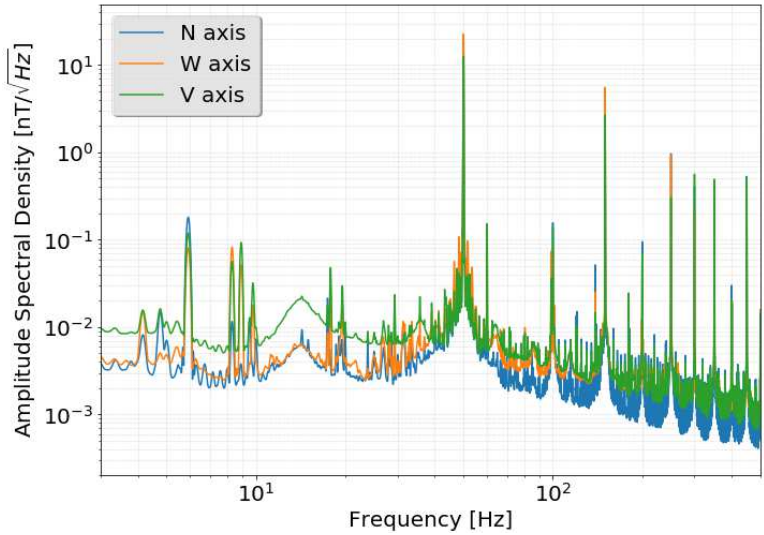}
\caption{Typical magnetic noise spectra detected at "Central Building" by the 3 permanent single-axis magnetometers used to monitor the magnetic activity in the experimental area. Periodic features and broader structures in the spectrum are easily identifiable as known sources. N, W and V stand respectively for North-South, West-East and Vertical directions.}
\label{fig:envB} 
\end{figure*}

There are plenty of different coupling mechanisms between a MN source and the interferometer; examples are a direct coupling to the magnets used for the TM control, but also via cables carrying signals to the voice-coil actuators to which the magnets belong.
Several countermeasures have been put in operation to reduce this kind of coupling (e.g. anti-aligned magnets, Shielded Twisted Pair cabling). However a residual signature is unavoidable, due to the extreme sensitivity of the interferometer. This kind of noise is very difficult to deal with, since the EM field can have a strong spatial dependency and it is usually hard to find any coherence between the strain signal and magnetic probes. When a coherence is found, the best practice is to act on the noise source, which means spotting its presence and location and eventually trying to cure it.
Typical sources are mains transformers and mains cables carrying high current, motors, impulsive electronics driving heavy loads (e.g. Pulse-Width Modulation regulators). Indeed any sufficiently close low power supply module and electronic board can also trigger a detectable disturbance. Other than that, many recurrent features are present in the AdV magnetic spectra (figure \ref{fig:envB}), ranging from a series of lines in the 4-10 Hz interval caused by the air conditioning system machinery to the 50 Hz and its harmonics, associated to the frequency of the electrical system.  
The principal lines due the the main power supply could be broadened by the presence of sidebands: they are barely discernible on the vertical magnetometer plot (V axis) in figure \ref{fig:envB} (e.g. around the 50Hz line, at 36 Hz and 64 Hz, and around the 100 Hz harmonic, at 86 Hz and 114 Hz). Sometimes, the sidebands are rather intense and can be visible as separate lines adjacent to the main one, causing undue coupling to the interferometer output and impairing the spectral sensitivity. 

\subsection{Evaluation of the vacuum chamber shielding properties}
\label{ssec:shield}

A study we carried out inside the WEB on February 2018 was a new evaluation of the magnetic filtering effect produced by the steel tank. The tank is a grounded austenitic stainless steel 304L shell with a diameter of 2 m and a thickness of 15 mm, which isolates the vacuum-sealed environment from the outside.
During a previous measurement (2006), a transfer function with frequency slope of $f^{-1.3}-f^{-2}$ was found, but there were high uncertainties on the shape of the filter and it was not possible to determine its exact order (first or second order) nor its pole (from about 5 to 30 Hz).

\begin{figure*}[ht]
\centering
\includegraphics[width=3in]{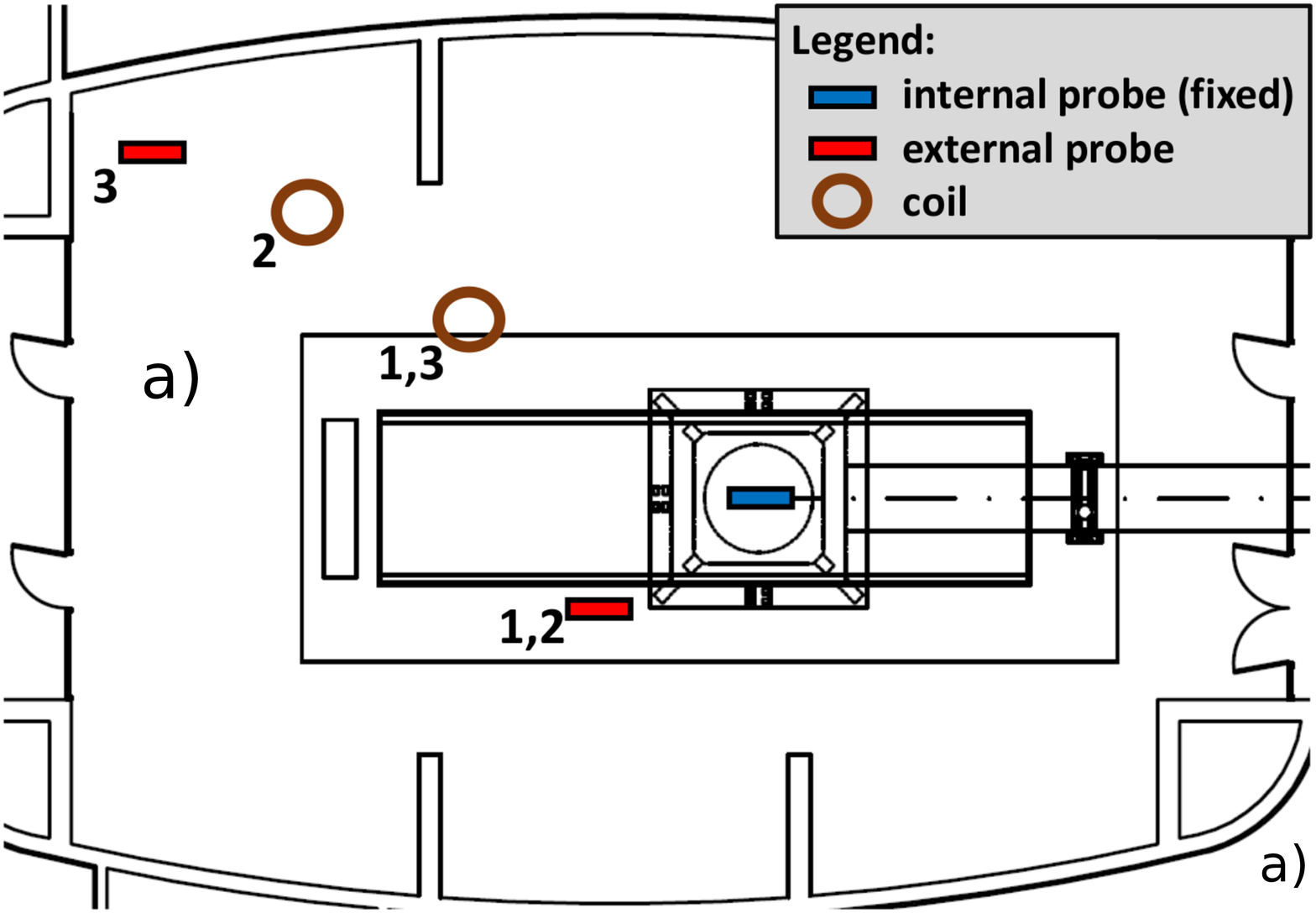}
\includegraphics[width=3in]{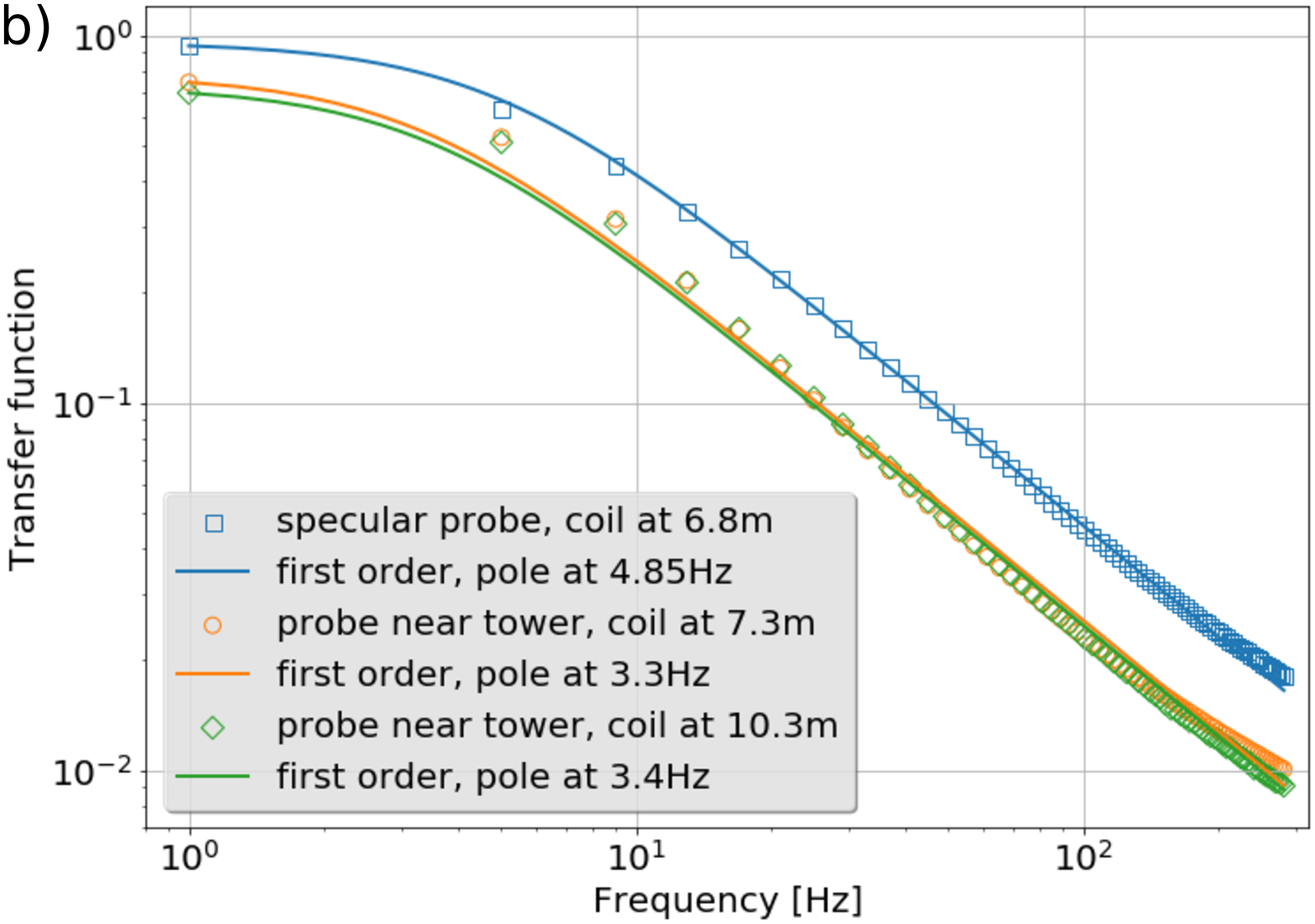}
\caption{Frequency evolution of the West-End Tower (WET) shielding power: a) The 3 "injection coil - external magnetic probe" configurations for their positioning inside the experimental area: coil at 7.3 m, probe near the WET (config. 1); coil at 10.3 m, probe near the WET (config. 2); coil at 6.8 m, specular probe (config. 3). b) The measurement results, along with the fits to a first order lower pass filter model of Butterworth type.}
\label{fig:Bshield} 
\end{figure*}

The new measurement required the positioning of an injection coil, which will act as magnetic field source, and two magnetic probes, one inside the Tower and the other outside.
Figure \ref{fig:Bshield}a shows a map of the different configurations: the 3 "injection coil - external magnetic probe" position pairs are numbered consecutively.
The shielding power of the steel enclosure is quantified by the filter frequency response $H(f)=B_{mod}^{inside}/B_{mod}^{outside}=k/\sqrt{1+(f/f_{pole})^2}$, where $B_{mod}$ is the magnetic field modulus, k is a constant and $f_{pole}$ is the frequency of the filter pole or in other words the cut-off frequency.
The trend of the experimental data fits the first order lower pass filter curve of Butterworth type with pole ranging from 3.3 to 4.85 Hz, depending on the setup configuration (figure \ref{fig:Bshield}b).

Above 100 Hz we note an inflection and a consequent change in the slope. This would mean the presence of an unknown effect, summing to that of a pure low pass filter (whose frequency response would indefinitely continue to decreasing). 

The difference between the measurements with the probe near the Tower and the measurement with the specular probe suggest that the magnetic field is more concentrated in the proximity of the Tower. Note that the frequency evolution is nearly the same for all the 3 configurations.

\subsection{Identification of point magnetic sources}

We know that ferromagnetic materials are usually magnetized during various machining phases. The causes of this residual magnetism may be different depending on the processing. Just to mention a few: forging, bending, welding, cutting, hardness checks, transport, handling and even mechanical vibrations. In our case, the main consequences we could encounter are a direct effect on the EM TM actuators and a potential disturbance on the instruments we use to probe the MN. As the NET can be mechanically excited by sound waves and seismic vibrations, magnetic patches on it could be a dangerous MN source. Hence we measured the DC magnetic field on the external walls of the NET to be sure there were no significant magnetization patches on it. We found an average field strength of about 35-50 $\mu$T in modulus on 24 locations around the tank, within a 1 m and a half height from the platform ground. These values correspond quite well to the average magnetic field measured in other internal and external building locations and to the typical magnetic field of the Earth. Therefore we don't see any evidence of alarming magnetic patches on the surface and the core of the steel tank at the height level of the TM.

\begin{figure*}[ht]
\centering
\includegraphics[width=4in]{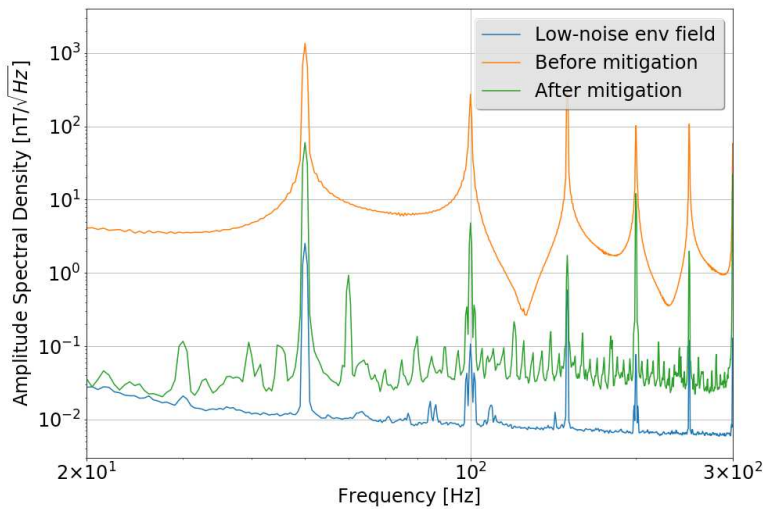}
\caption{Comparison between three different magnetic noise levels, found near a localized noisy source, before (orange) and after its removal (green) and in correspondence of a particularly low-noise location (blue), inside the "North-End Building".}
\label{fig:EMactivity} 
\end{figure*}

Nevertheless, there may be other higher MN sources around the tank. We found a particularly high and variable (in intensity and periodicity) EM activity in the proximity of the electronic racks, which are almost two metres away from the NET. However the intensity of this disturbance drops to the standard quiet magnetic activity ($[0.01,0.1]$ nT/$\sqrt{Hz}$) before reaching the tank.
Another example of noise coming from fields irradiated by electronic devices was tracked down to power supply boxes, whose noise was nearly 3 orders of magnitude higher than the standard quiet magnetic activity. These boxes were placed close to each Tower, on the first floor platform. We performed some on-off tests and then we removed the noisy power supply.
In figure \ref{fig:EMactivity} we compare the noise level near the localized MN source, before (orange) and after its removal (green), with the environmental low-noise activity in the area (blue). This preliminary characterization and mitigation work was necessary for the evaluation of the ambient noise contribution to the AdV sensitivity.

\subsection{Far-field magnetic noise injections}

Stimulated noise injections are largely used in Virgo to assess to what extent a noise would spoil the instrument sensitivity. This approach is called ‘active’ as we stimulate the interferometer (or just small sections of it) with an active probe in order to drive a discernible signal in the instrument output.

In order to study the MN, we had to build from scratch a driving coil. The coil is composed of a 1 mm section copper wire, wrapped 50 times to form a 1 m diameter winding around a full PVC frame as a support structure.

We extract the magnitude of the environmental or injected magnetic fields with two different kind of probes. Firstly the single axis, one meter long, tube-like magnetometer MFS-06 by Metronix, with a very low intrinsic noise (better than 0.01 pT$/\sqrt{Hz}$ at 10 Hz), which makes it suited to measure the field strength variations of the Earth. Secondly a portable triaxial magnetic field sensor FL3-100 by Stefan Mayer Instruments with a measurement range of $\pm 100\ \mu$T and an intrinsic noise of few pT$/\sqrt{Hz}$ at 10 Hz.

A very crucial aspect to take care of is the positioning of the coil and the magnetic probes inside the experimental buildings. More specifically we want the interferometer and the magnetometers to sense approximately the same injected field in order to obtain a faithful projection. Moreover we want to be in the regime of the far-field injections, where the source is ideally situated at infinite distance from the detection equipment (coil dimensions << distance). If we take into account the space limitations of the area surrounding the vacuum chambers, we were left with limited choices to the positioning of the instrumentation. A quick calculation shows that, with a current of about 10 A flowing in the coil, the magnetic field strength at 20 m from the coil would be around a few tens of $\mu$T.

\begin{figure*}[ht]
\centering
\includegraphics[width=6.1in]{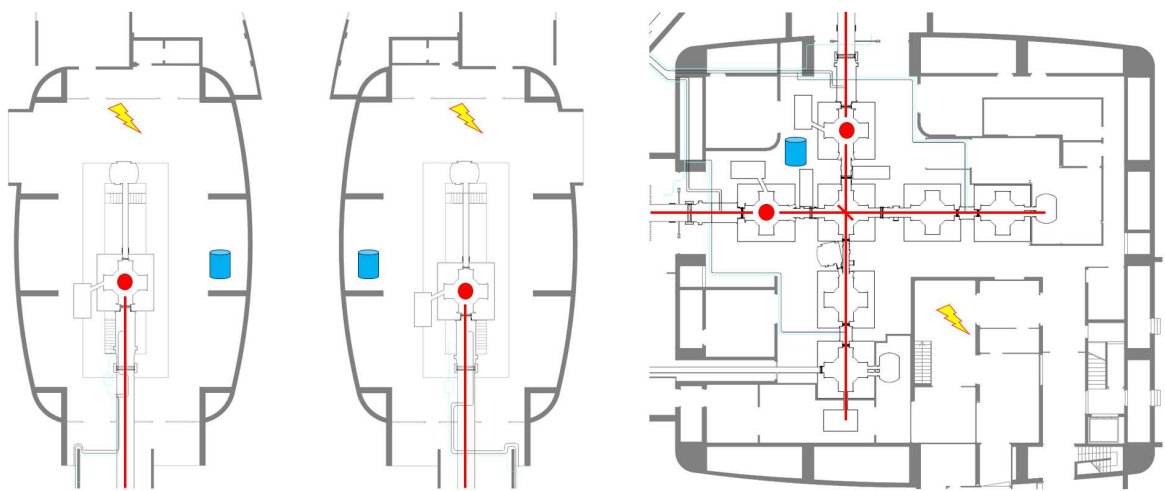}
\caption{From left to right the "West-End, North-End and Central Buildings" floor plans. The thunderbolt and the blue cylinder correspond respectively to the coil and the probe locations inside the three facilities; the red spots identifies the four TMs and the red lines follow the main laser path.}
\label{fig:coilprobe_pos} 
\end{figure*}

The first measurement campaign was performed during the second half of 2017 inside the three main AdV building: NEB, WEB and CEB. The same procedure has been replicated every time by adapting the setup to the building environment. The three actual coil-probe mutual positions are depicted in figure \ref{fig:coilprobe_pos}.

\begin{figure*}[ht]
\centering
\includegraphics[width=3in]{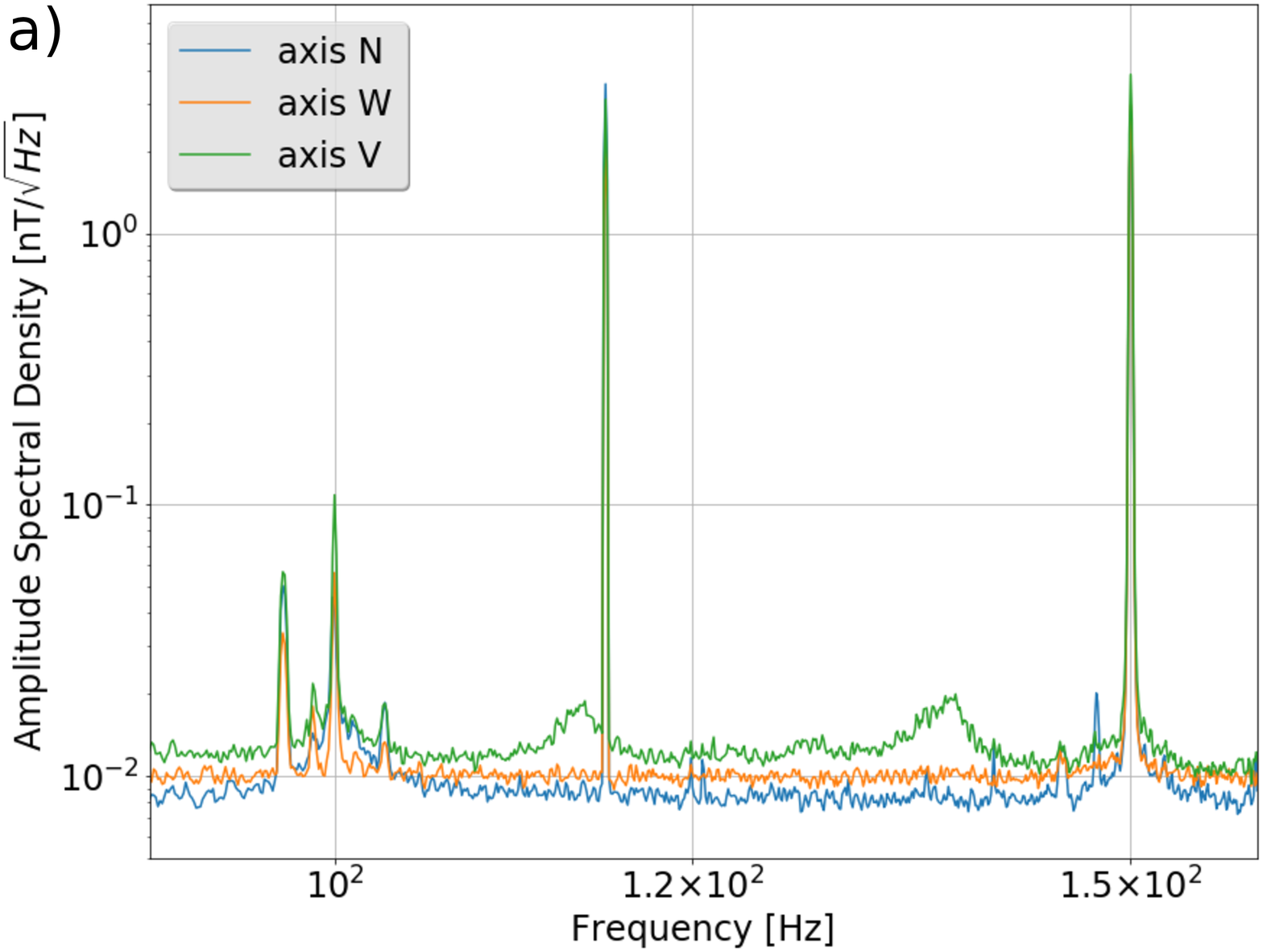}
\includegraphics[width=3in]{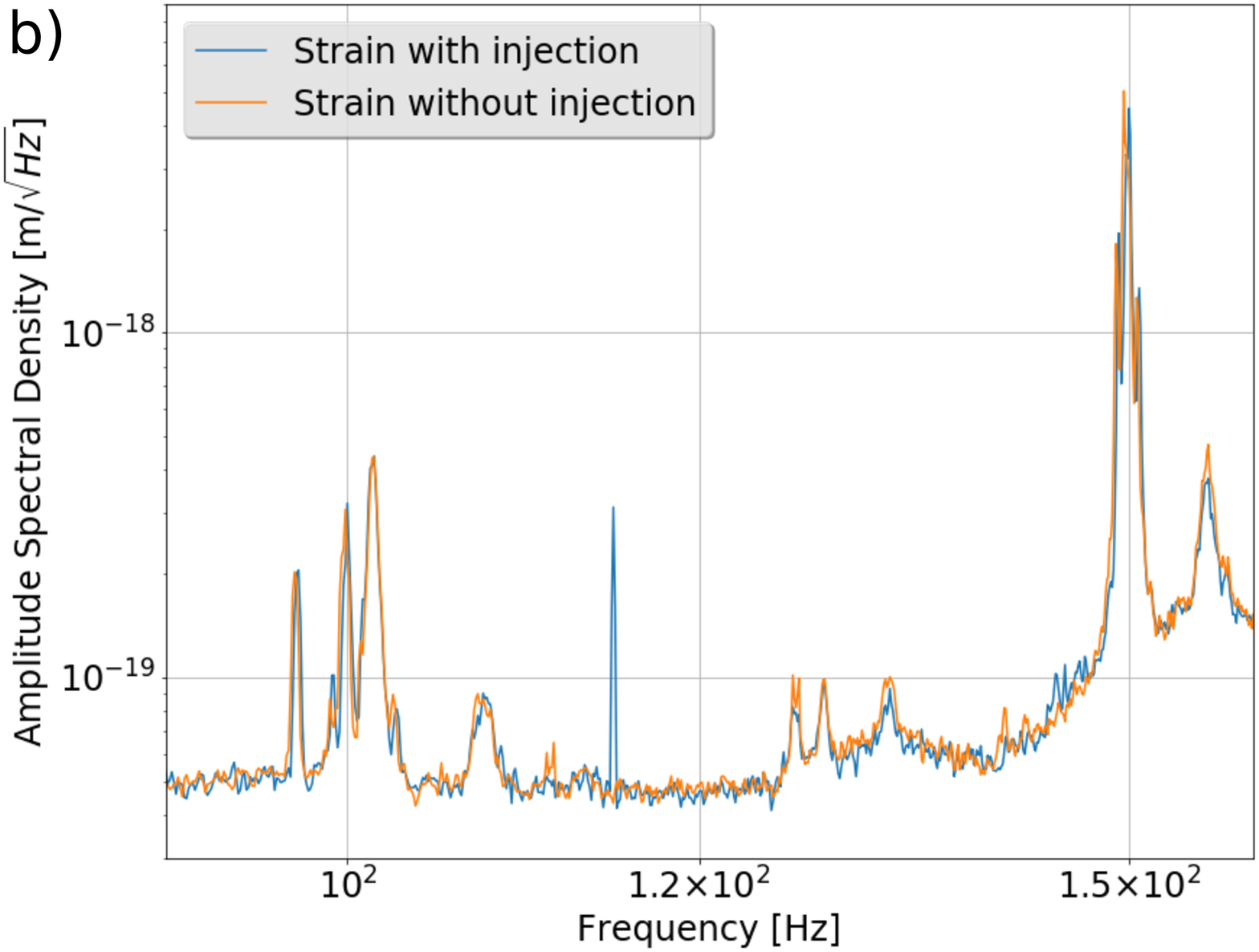}
\caption{114.8 Hz line injection at "Central Building". It is readily visible by the 3 single-axis magnetic probes (a) and the interferometer (b), where the blue peak stands out against the reference orange curve.}
\label{fig:maginj}
\end{figure*}

We injected a series of monochromatic magnetic signals (magnetic lines) lasting 600 seconds each and with field intensity of few mT at the magnetometer locations for each line and for the entire explored frequency range (from 14 to 140 Hz). For comparison, this is nearly the same field strength we encountered in correspondence of the noisy source of figure \ref{fig:EMactivity} below 100 Hz.
At the same time we need some important conditions to be met during the procedure: interferometer in science mode\footnote{Fully locked interferometer in the foreseen configuration for observation, with no human or automated intervention on the machine.}; quiet environmental magnetic activity\footnote{Absence of known noisy magnetic field sources close to the TM, for instance due to accessory/monitoring electronics and/or during a thunderstorm.}; sensors readout far from saturation; the amplifier tuned so that the output measured on the spectral density had the same amplitude for all injected lines, regardless of the input frequency (from a few Hz to a few hundred Hz); good coherence between the injected lines and the strain signal h(t) for almost every investigated frequency. In figure \ref{fig:maginj} we show the injection of the 114.8 Hz line at CEB, which is visible both by the magnetic sensors (\ref{fig:maginj}a) and the detector (\ref{fig:maginj}b). The overall coherence is above 0.8 for almost all the other frequencies, except for the 14 Hz due to poorer detector sensitivity.

\begin{figure*}[ht]
\centering
\includegraphics[width=4in]{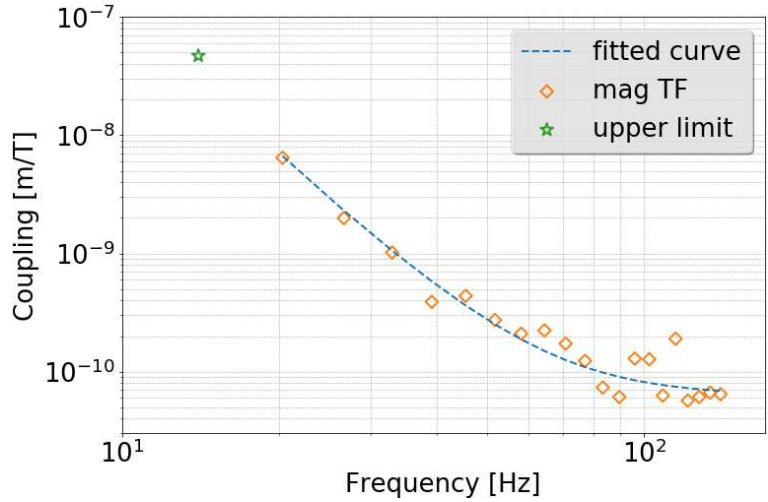}
\caption{Magnetic coupling measurement. The orange diamond means that the corresponding injected line is detected in the strain channel, while the green star is only an upper limit, due to bad coherence between the injected field and the strain output. The fit to the coupling function is also included.}
\label{fig:TF_fit} 
\end{figure*}

We repeated the injection scan several times to make sure we have always stable interferometer conditions.
After that, the responses of the sensors and the instrument are analyzed, in order to quantify the coupling between the environmental MN (m) and the strain (h). The transfer function, at the time of the injections, corresponds to:

\begin{equation}
TF_{mh}(f)=\frac{ASD_{hm}^{\ inj}(f)}{ASD_{mm}^{\ inj}(f)}
\label{eq:tf_inj}
\end{equation}

where ASD stands for Amplitude Spectral Density (cross spectral density as a numerator and power spectral density as a denominator). We computed the coupling for all measurements inside the 3 buildings and took the incoherent sum to obtain a single magnetic TF, which is displayed in figure \ref{fig:TF_fit}. We fit the coupling measurements with a power law and performed a least squares fit of the model to the data to estimate the model parameters. The results show that the points follow the model below:

\begin{equation}
TF_{mh}(f)=TF_{mh}\left(\frac{f-f_{0}}{10\ Hz}\right)^{\alpha}+\beta
\label{eq:tf_fit_model}
\end{equation}

with amplitude $TF_{mh}$ and pole $f_{0}$. Most notably, the exponent $\alpha=-3.3\pm0.4$ is consistent with the expectations regarding the decreasing trend of the coupling function. Indeed we expected a factor -1 from the tank shielding, plus a factor -2 from the TM pendulum mechanical contribution. Above 100 Hz, spurious phenomena like the electronic coupling through cabling are probably dominant and add up to the descending pattern given by the low-pass filtering of the Tower.

For what it concerns the coherence between the injected lines and h(t), we considered the frequency range $[20,140]$ Hz reliable, while at lower and higher frequencies the lines are not always perfectly visible by the interferometer and this could lead to an overestimate of the projection.
However we also investigated higher frequency coupling, whose results can be found in section \ref{sec:results}, in terms of comprehensive magnetic noise projections.

Indeed the last step involves the computation of the magnetic projection by multiplying the coupling function by the environmental magnetic spectrum (no injections):

\begin{equation}
P(f)=TF_{mh}(f)\cdot ASD_{mm}^{\ quiet}(f).
\label{eq:proj_inj}
\end{equation}

The values from equation \ref{eq:proj_inj} work under the assumption that both sensors and the Tower experience the same field intensity, which is true in the far-field regime. However, in our case the distance between the driving coil and the Tower is hardly greater than 20 m and hence the positioning of the magnetic sensors becomes a crucial aspect. We verified that the transfer function slope is conserved even when we changed the location of the magnetic sensors. In our test, we fixed the coil location and placed three different set of magnetic sensors, including one nearer to the coil and another one farther away from the coil with respect to the Tower. While the transfer function slope is conserved, there was an amplitude shift so that we obtained the amplitude range on the projected noise (the nearest sensor underestimates the intensity, the farthest one overestimates it).

In addition, we assessed the MN effect in different buildings. It turned out that the coupling level at CEB is comparable to the one we measured at the WEB and NEB buildings. In a later measurement campaign, we were able to repeat the injections at CEB with the optimized AdV sensitivity, which we confidently take as a proxy of the other two locations (NEB and WEB).

\subsection{Magnetic noise projection with coherence}

The far-field projections and the simulation-based approach are not the only ways to estimate the noise contribution to the sensitivity: a third possibility is to use the coherence between the magnetic and the GW channels as if the background magnetic activity were the (external) injected field. The main limitation is that we need a long time interval in order to get a meaningful value of the coherence, as the MN does not contribute significantly to the GW signal. With this latter method, we should obtain an upper limit of the MN level, which we can compare with the projected noise contribution we already got from the injections.

The background coherence analysis lasted for 10 hours, as data were acquired during a stable functioning interval of the interferometer (i.e. locked). The starting time point was on Monday 08 Apr 2019, at 12:00:00 UTC, and during the data gathering period the Binary Neutron Star observing range was around 50 Mpc. These figures were sufficient to calculate the coherence under stationary conditions by averaging over 360 data chunks, each one 100 seconds long (50$\%$ overlapping).

\begin{figure*}[ht]
\centering
\includegraphics[width=3in]{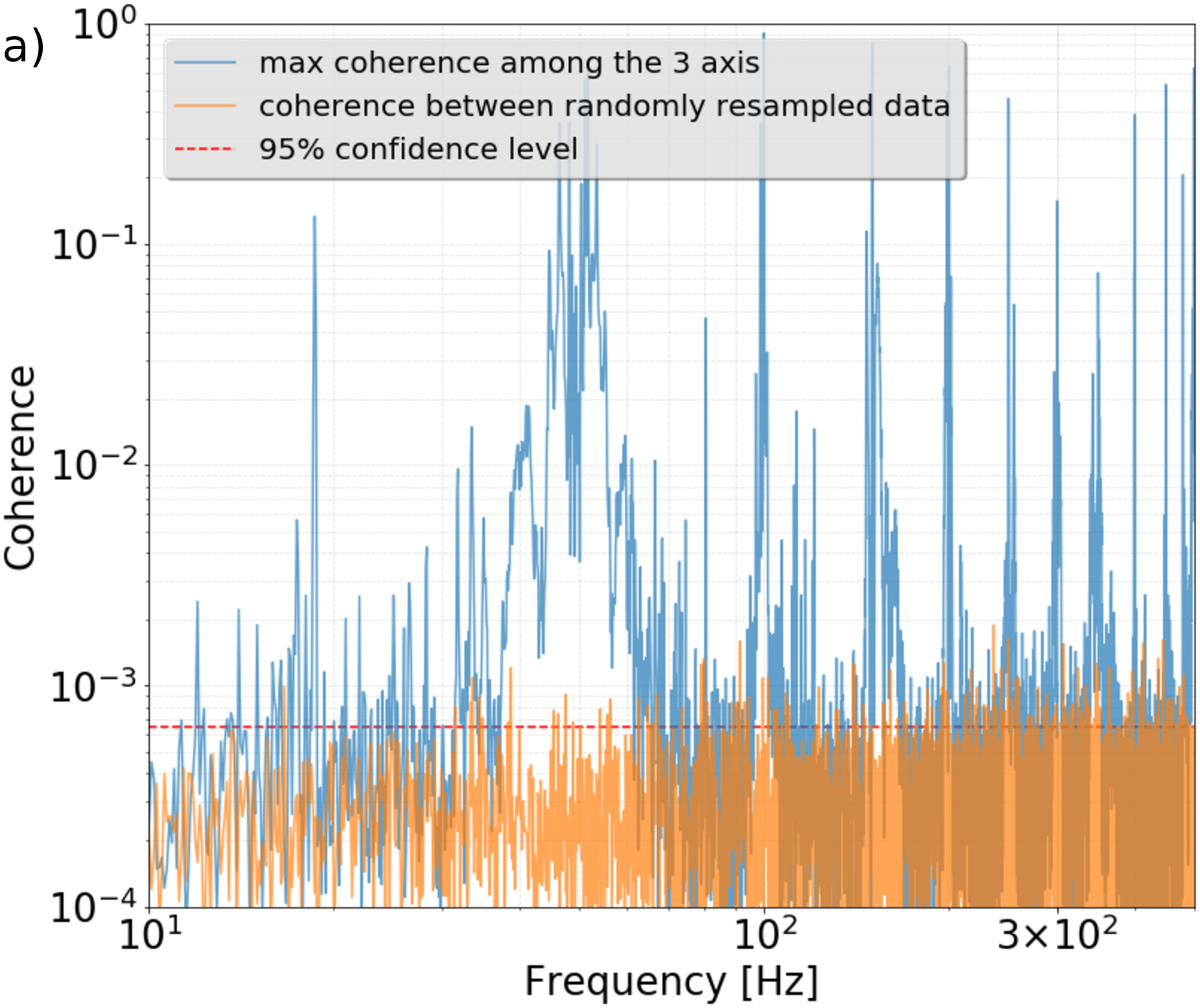}
\includegraphics[width=3in]{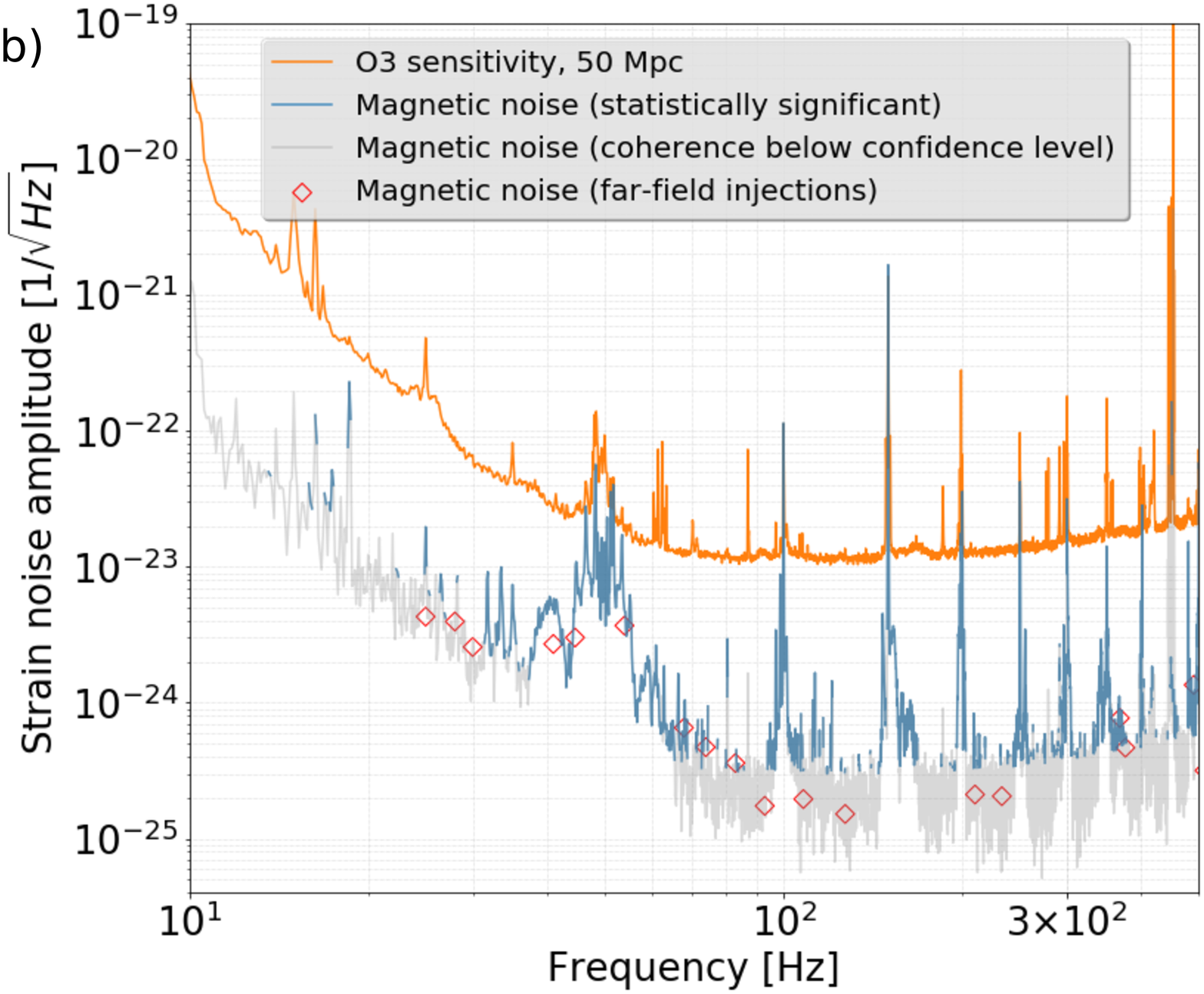}
\caption{Magnetic noise projection with coherence. a) Maximum coherence at each frequency between the strain channel and the 3 magnetic probes in the "Central Building" (blue line); the dashed red line indicates the statistical significance level of the coherence, calculated according to the 95$\%$ confidence level of the coherence distribution between randomly resampled data (orange line). b) Related magnetic noise projection (blue for statistically reliable bins, grey for coherence below significance level), compared to the most recent one obtained by the far-field injections (red diamonds) and the Advanced Virgo sensitivity (orange line).}
\label{fig:cohere} 
\end{figure*}

Figure \ref{fig:cohere}a shows the max coherence value among the GW output channel h(t) and the 3 magnetometers in the CEB (blue line). The statistical significance of the computed coherence (dashed red line) is the 95$\%$ confidence level of the coherence distribution between randomly resampled data (orange line). Figure \ref{fig:cohere}b shows a noise projection calculated according to:

\begin{equation}
P_{cohe} = \sqrt{C_{mh}}\cdot ASD_{hh}
\label{eq:proj_cohe}
\end{equation}

Here the statistically reliable coherence bins are drawn in blue, while the remaining ones in grey.
Results from the background coherence analysis are in good agreement with the ones obtained by the far-field injections acquired in the CEB during the latest measurement campaign (red diamonds). Incidentally, these results come from the analysis of the same time period. The advantage of the background coherence method is that we obtain noise projection over all frequencies, as opposed to the discrete sampling of the frequency interval in the case of the artificial injections. The drawbacks are the need for long time data series in order to get an acceptable level of coherence and the influence of glitches that might occur during this particularly long time period. For this reason we vetoed data sections using a glitch monitoring channel. Indeed, it could be interesting in the future to further investigate the broadening of the coherence around 50 Hz. For now, we can only speculate on the origin of these structures: possible causes could stem from significant phase-noise fluctuations of the mains (we remind the acquisition time of 10 hours) and/or non-linear behavior of the various power supplies and electronics.

\section{Results}
\label{sec:results}

\begin{figure*}[ht]
\centering
\includegraphics[width=5.5in]{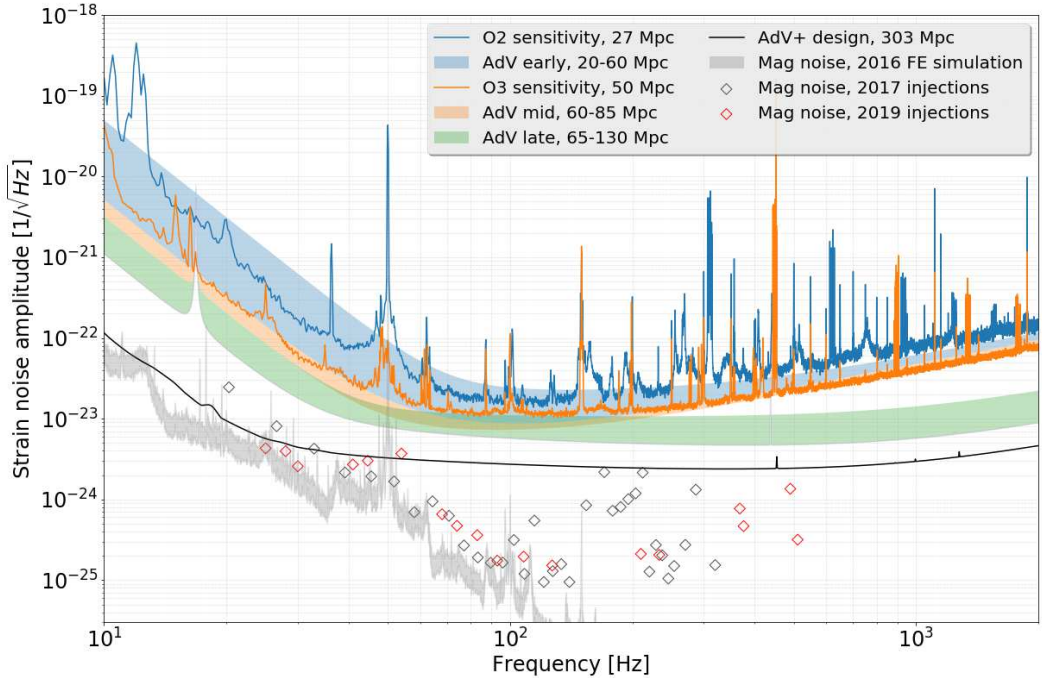}
\caption{Magnetic noise budget. The relevant Advanced Virgo and Advanced Virgo+ present and future sensitivity outcomes are shown as reference. The grey band represents the "Finite Element" simulation uncertainty among all the investigated electrical configurations of the payload, which is the mechanical assembly that suspend the test masses. Diamonds represent the far-field magnetic noise injections, performed 2 years apart.}
\label{fig:allproj} 
\end{figure*}

Finally, we computed the magnetic noise budget. Figure \ref{fig:allproj} shows a comprehensive view of the AdV and AdV+ observing scenarios and the MN estimates. We specifically compare two completely independent evaluations: the FE analysis done in 2016 \cite{cirone_magnetic_2018} and the experimental noise injections (both the 2017 and 2019 ones). The FE simulation study took into account a large number of possible electrical configurations of the payloads and eventually the sum of translational and rotational effects were taken. This simulation considers the impact on the 4 mirrors of the 2 Fabry-P\`{e}rot cavities. Hence the comparison with the magnetic injections was made possible as we incoherently summed in quadrature the injections performed in the three main AdV buildings. The FE analysis provides a slightly more optimistic estimate than the one obtained with the current study but, on the other hand, the slope of the strain noise spectrum (below 100 Hz) appears to be the same around $f^{-3}$. We also point out the substantial consistency of the MN level during a 2-year period, from 2017 to 2019. At frequencies higher than 100 Hz, simulation and injections don't agree anymore, with the former continuously decreasing and the latter spreadly flattening. The rationale for this behavior is currently under investigation but it seems plausible to attribute it to the magnetic coupling with nearby electrical cables, an effect not taken into account in the FE simulations. 
However the influence of the MN on the instrument sensitivity is not concerning for the O3 observing run yet, but it might be for the next interferometer design phases.
A similar study of the magnetic coupling was carried out at the advanced Laser Interferometer Gravitational-Wave Observatory (aLIGO) \cite{advLIGO}, with comparable results \cite{ligo_mag} (see the "Discussion" section for further details).

\section{Proposal for a mitigation strategy}

As a final step, we investigated a mitigation strategy to reduce the MN to avoid limitations even at the future design sensitivities. For this purpose we studied different passive shielding configurations consisting of Helmholtz coils and/or spherical metallic grids to be put around each TM vacuum chamber. 
The idea is to passively reduce the field intensity in the area close to the mirrors in order to reduce gradients at the magnets position. This works for noisy sources at both the external and internal sides of the shield.

\begin{figure*}[ht]
\centering
\includegraphics[width=1.45in]{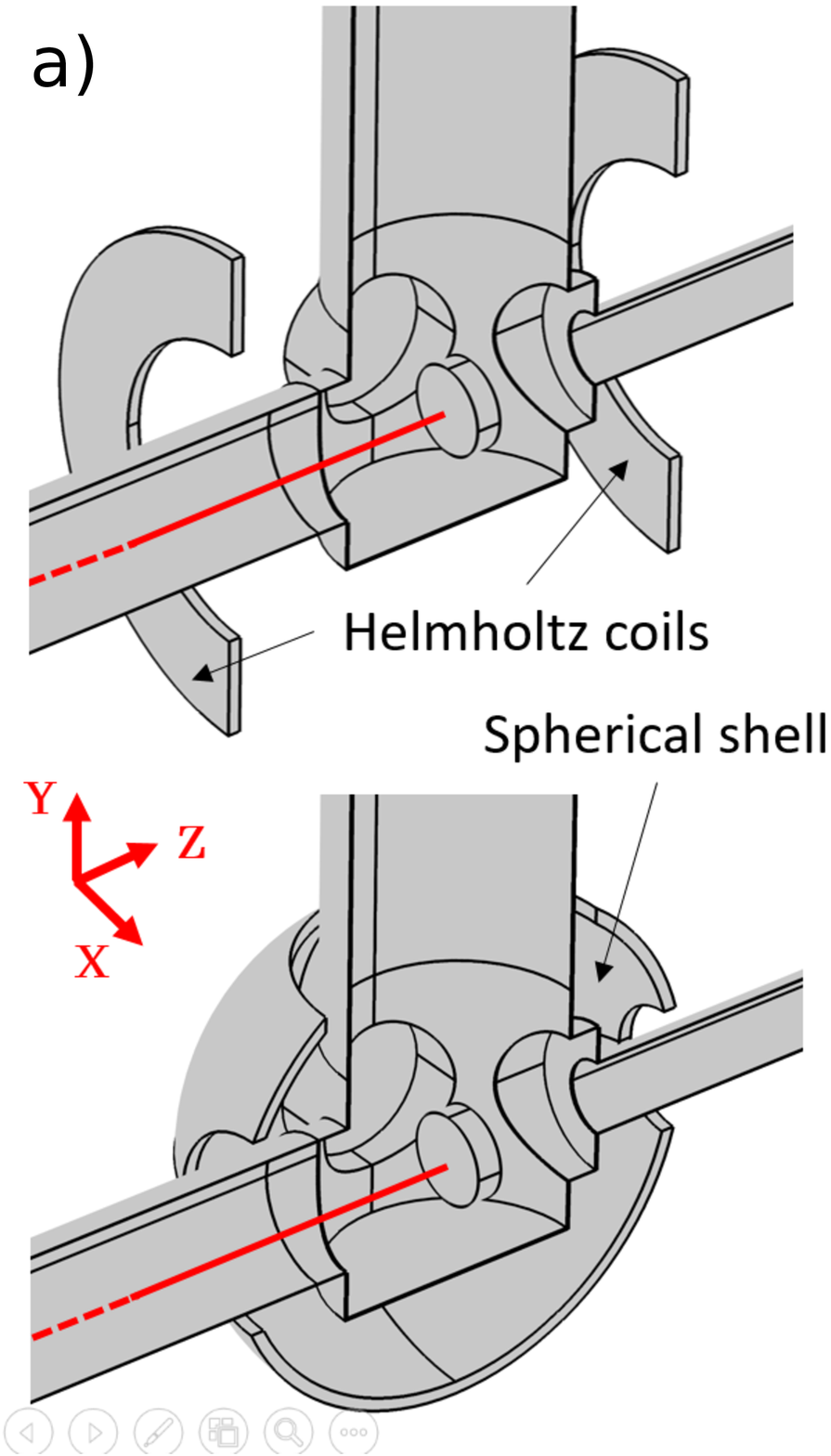}
\includegraphics[width=3.8in]{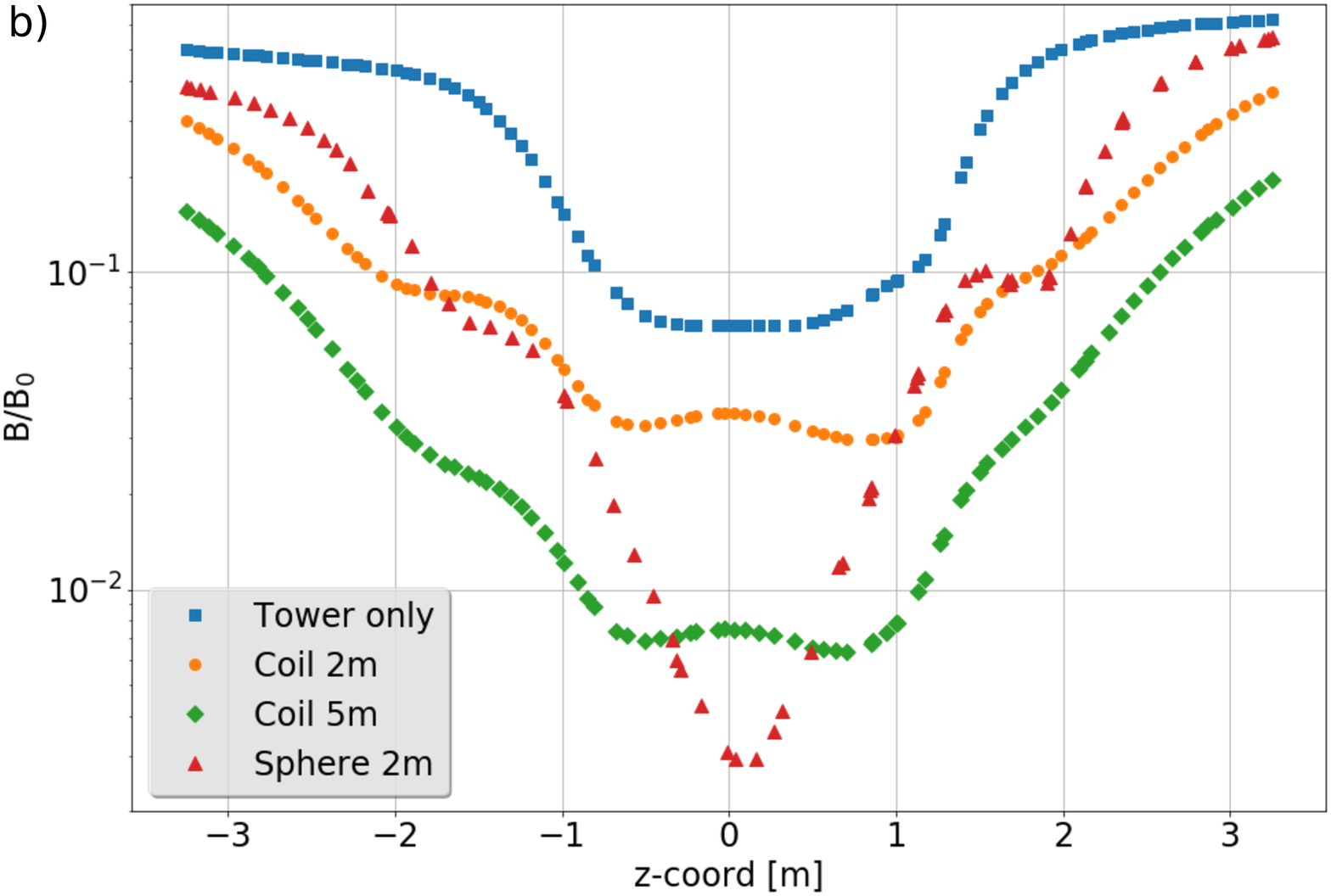}
\includegraphics[width=3in]{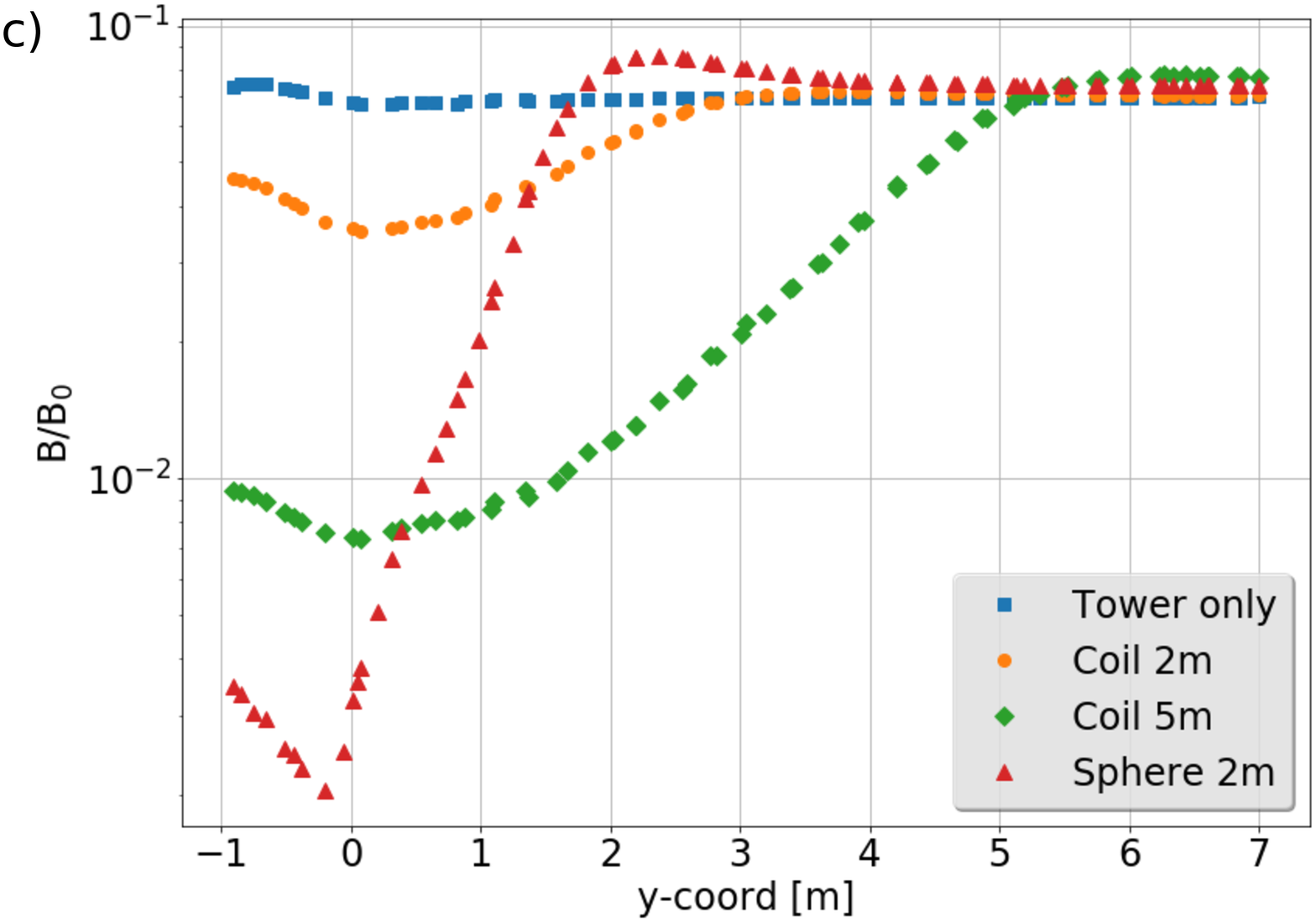}
\includegraphics[width=3in]{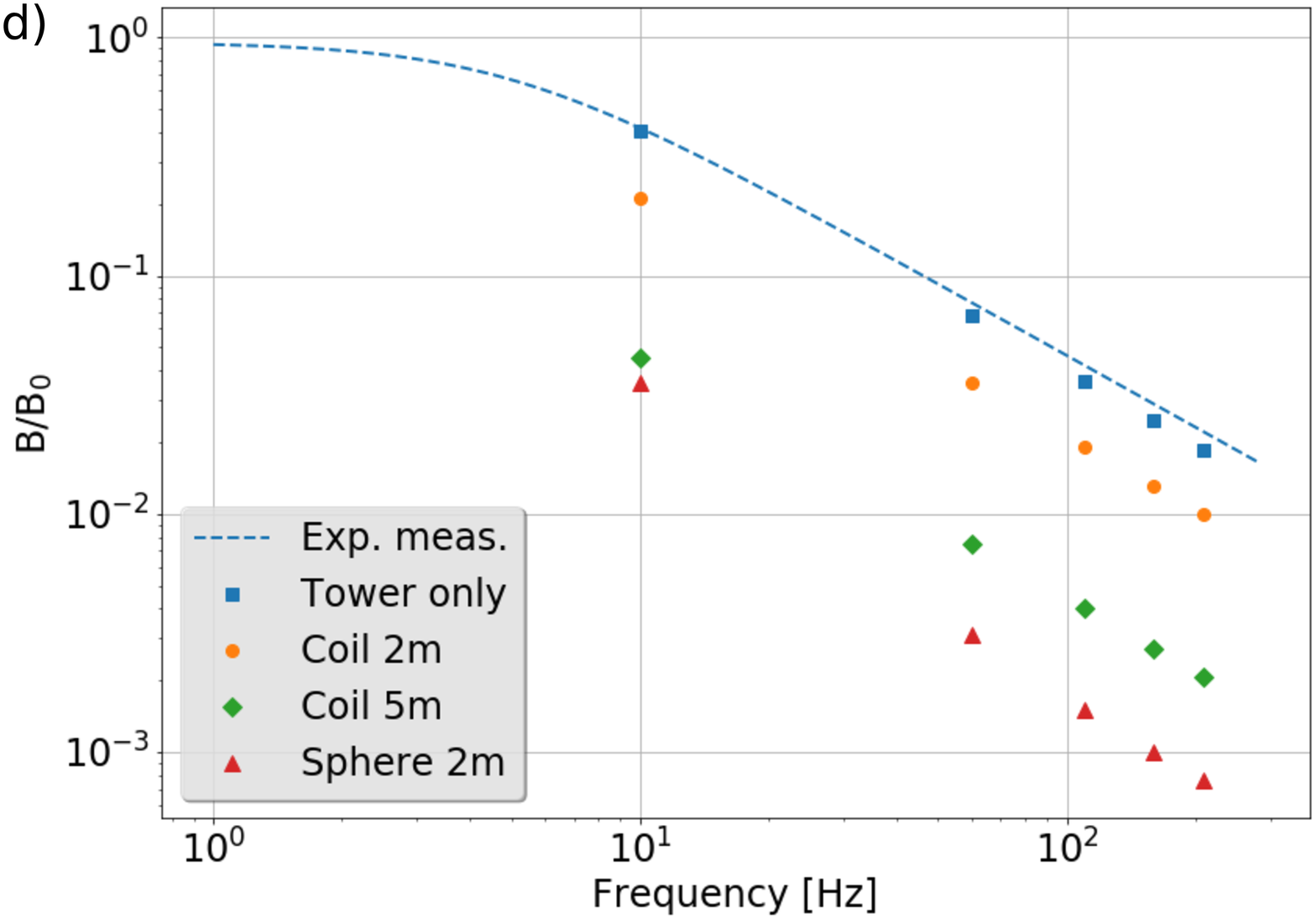}
\caption{"Finite Element Analysis" of two possible magnetic field shielding solutions: a) the Helmholtz coils and a spherical shell (different shapes and sizes). b) Longitudinal studies (along z) of the normalized reduction of the magnetic field. c) Orthogonal studies (along y) within a height range of [-1,7] m, along the "Tower" axis. d) Frequency evolution of the normalized field at the Test Mass location.}
\label{fig:mitig} 
\end{figure*}

We used the Finite Element Analysis software "Comsol Multiphysics" to simulate the coarse geometry and physics. We recreated the vacuum chamber surrounding the TMs with the exact same dimensions of the real one and tried several different shielding configurations. We report results on the best two solutions: the Helmholtz coils and a spherical shielding made of chain-link (figure \ref{fig:mitig}a). The former have an external radius of 2m/5m, an internal radius of 1.2m, a thickness of 0.1m and they are placed 2m far from the mirror. By contrast the metallic sphere is centered on the TM position, it has an external radius of 2m and a thickness of 0.1m. The FE model was consequently exposed to a spatially homogeneous and harmonic magnetic field of intensity 1T along z.
It turned out that Helmholtz coils of radius 5m are able to lower the external magnetic field up to a factor 10, with respect to the intrinsic Tower shielding, and homogenize it within 1m from the centre of the mirror both along y and z axis (figure \ref{fig:mitig}b-c). In addition, the shielding effect (evaluated at the virtual TM location) increase with the frequency (figure \ref{fig:mitig}d) and the coil turns, as expected. 

On the other hand, the spherical grid configuration reached a better level of reduction up to a factor 20 but in a narrower region. This point is supported by both the longitudinal and orthogonal profiles. In this sense we feel confident to consider the integration of the Helmholtz coils configuration as the better possible future mitigation strategy, essentially in terms of improved homogenisation power, easier implementing and less overall space required. As further validation, we got a good agreement between the Tower shielding experimental measurements (with the filter pole at 4.85 Hz, see section \ref{ssec:shield}) and the Tower numerical simulations (without any further shielding, see figure \ref{fig:mitig}d).

This was an exploratory mitigation analysis that we may further pursue in order to be compliant to the sensitivity targets of the detector designs. Of course, separate strategies might be taken into account as well, like for example a new TM actuation method that do not require any parts on the TMs with magnetic susceptibility. MN mitigation is going to be important also for the upcoming realization of the Einstein Telescope, as its spectral sensitivity would be at least one order of magnitude higher than the AdV design one. 

\section{Discussion}

In this paper, we have studied the MN contribution to the AdV sensitivity.
Our approach was experimental, and it was compared to simulation studies of various EM phenomena and their impact on the interferometer configuration. 
For this study, we had to deepen our understanding of the various local EM sources and their effects on the measuring process.
We started with the evaluation of the EM shielding properties of the steel tanks embedding the TMs. The tank mostly behaves as a low-pass filter as expected, albeit with a mild slope inflection above 100 Hz.
Then we identified and mitigated whenever possible the most annoying EM sources nearby the TM Towers, including the verification of the presence of any residual magnetism on the metallic enclosure walls. Indeed, a MN source like that of figure \ref{fig:EMactivity} would affect the AdV sensitivity below 100 Hz, if not removed.

With this preliminary work we prepared the environment for the MN injection campaign, in order to directly assess the ambient MN effect on the AdV output. Results show that there is substantial agreement between simulations and noise injections, and that MN is currently not an issue for the AdV sensitivity.
There are, however, some discrepancies between the simulation and the noise injections at frequencies > 100 Hz, which suggest the presence of additional effects. The slope flattening in the transfer function is real (see section \ref{ssec:shield}) and it can't be ascribed to the shielding properties of the metallic vacuum chamber. A possible explanation is that the simulation is limited to the Payload structure and does not consider all the extra components, passive or active, in the area. 
Interestingly aLIGO noticed a very similar behaviour, therefore we think that it can be attributed to the magnetic coupling to all the electronic cables in the proximity of the vacuum chambers. 
While aLIGO uses a very different actuation scheme, with mixed electrostatic and EM actuators \cite{advLIGO}, AdV uses exclusively EM actuators for all the positioning corrections. Accordingly, AdV is expected to be more plagued by any magnetic disturbance. As a result, aLIGO is probably more dominated by coupling to cables and electronics, cable connectors and so on, which could be influenced by cabling issues and could enter the noise path through the marionette actuation stage. In this case, the magnetic noise impact should drastically change along with the general maintenance and commissioning of the interferometer on short time scales. For further reference, see the aLIGO technical note \cite{magTF_ligo}.

The small discrepancy found below 100 Hz between the payload magnetic simulations and the magnetic injections could also be explained by the updated payload design, whose details were not included in the previous study and cannot be easily accounted for due to their small details and intricacies (for instance the installation of Fiber Guards to shield fused silica fiber from high speed particles \cite{payload_FG}).

An additional issue related to magnetic coupling is the correlated MN from Schumann resonances, which is relevant to the observation of a Stochastic Gravitational-Wave Background (SGWB) in GW detectors \cite{thrane2013,schumann_mio}. Briefly, Schumann resonances are global EM signals in the cavity formed by the surface of the Earth and the ionosphere. The cavity is excited by the lightning activity around the world and eventually a magnetic field of $0.5-1.0\ pT/\sqrt{Hz}$ is produced on the Earth's surface, which is roughly one order of magnitude lower than the environmental MN at CEB (see figure \ref{fig:envB}). Correlated noise, however, cannot be reduced through integration nor can it be mitigated through instrumental (re)design and/or background subtraction. For this reason a LIGO-Virgo joint experimental evaluation of the global MN contribution may be worth investigating in the future. In anticipation of that, it has been decided to keep monitoring the global MN activity at each GW site with dedicated low-noise magnetometers \cite{schumann_mio}. Similarly it has been proposed to perform MN injections on a weekly/monthly basis, during the normal scientific run, to keep track of the magnetic transfer function. The injections would be made with a series of permanent coils at each building, lasting less than an hour per injection cycle.

In the next few months, aLIGO and AdV will complete the third observing run with increasingly large sensitivity and number of detections. The development of noise reduction techniques to improve the opportunities of detection, especially for low SNR events, is therefore a crucial effort. Our mitigation strategy, consisting in the integration of a passive Helmholtz coil system around the most sensible TM Towers, may thus prove to be a valuable one.
The present study has shown that, if the GW detectors reach their design sensitivity, some MN mitigation strategy might become relevant. That is also the case for the future third generation ground-based GW detectors as the European Einstein Telescope (ET) \cite{ET_2011}, whose low frequency spectral sensitivity would be certainly limited by the above documented MN level.

\ack
The authors gratefully acknowledge the European Gravitational Observatory (EGO) and the Virgo Collaboration for providing access to the facilities and to the environmental data. We would particularly like to thank Antonio Pasqualetti for providing the CAD drawings of the Virgo experimental areas. AC was supported by the INFN Doctoral Fellowship at the University of Genova. The member of IFAE-Barcelona group were supported by  the Spanish Ministry for Science, Innovation and Universities, and the Catalan CERCA Programme.

\section*{References}
\bibliographystyle{iopart-num}
\bibliography{references.bib}
\end{document}